\documentclass{pasj00}

\begin{document}
\title{Photometric Observation and Numerical Simulation of Early Superhumps in BC UMa during the 2003 Superoutburst}

\author{Hiroyuki \textsc{Maehara}, \textsc{Izumi Hachisu}}
\affil{Department of General Systems Studies, Graduate School of Arts and Sciences,\\ University of Tokyo, 3-8-1 Komaba, Meguro, Tokyo 153-8902, Japan}
\email{maehara@provence.c.u-tokyo.ac.jp, hachisu@chianti.c.u-tokyo.ac.jp}

\author{Kazuhiro \textsc{Nakajima}}
\affil{Variable Star Observers League in Japan (VSOLJ), 124 Isatotyo Teradani, Kumano, Mie, Japan}
\email{K.Nakajima@ztv.ne.jp}

\KeyWords{
accretion, accretion disks
---
stars: dwarf novae
---
stars: individual (BC UMa)
---
stars: novae, cataclysmic variables
} 

\maketitle
\begin{abstract}
We carried out time-resolved photometric observations of an SU UMa type dwarf nova, BC UMa, during its superoutburst in February 2003.
We detect early superhumps ( or outburst orbital humps) during the first two days 
of the outburst. 
The period of early superhumps is 0.06258(13) d and the amplitude is 0.04 mag. 
After the early superhump phase, common superhumps with an amplitude of 0.3 mag and a period of 0.064466(16) d developed.
The change rate of superhump periods ($\dot{P_{\rm sh}}/P_{\rm sh}$) was positive through the superoutburst.
The superhump period excess ($\varepsilon \equiv P_{\rm sh}/P_{\rm orb} - 1$) is 3\% and we derive a mass ratio of 0.13.
This is twice as large as that of WZ Sge,
suggesting that the mechanism of early superhumps in BC UMa
 is not the 2:1 resonance which was proposed in WZ Sge.

We have modelled early superhump light curves including irradiation
effects of the accretion disk and secondary star by the white dwarf and accretion disk.
The observed early superhumps can be reproduced when two-armed spirals appear
 on the accretion disk.

We have found the long term data taken from AAVSO, VSOLJ and VSNET shows BC UMa has
normal outbursts with a maximum magnitude of $m_V\sim 13$ and 
two types of superoutbursts: one has a short duration (around 10 days) and 
faint maximum magnitude ($m_V\sim 12.5$), the other has a long duration
(around 20 days) and bright maximum ($m_V\sim 11-11.5$).
BC UMa is the first example of dwarf novae showing the two types of superoutbursts.
The supercycle of BC UMa is between 600 days and 1000 days. This is
shorter than WZ Sge type dwarf novae and longer than normal SU UMa type
stars.
This phenomenon suggests that BC UMa is an intermediate dwarf nova
between WZ Sge and SU UMa.
\end{abstract}

\section{Introduction}
Dwarf novae are a class of cataclysmic variables (CVs), which are close
binary systems consisting of a white dwarf and a red dwarf or a main sequence
star. SU UMa type dwarf novae are a subgroup of dwarf novae characterized by
 having two types of outbursts: 
normal outbursts (short duration and faint maximum brightness) and
superoutbursts (long and bright maximum) (e.g. \cite{Warner1985}).
A unique property of superoutbursts is the appearance of superhumps.
Superhumps are small-amplitude periodic modulations that are observed only
during the superoutburst.
Superhump periods are a few percent longer than the orbital period.

WZ Sge type dwarf novae are a subtype of SU UMa type,
properties of which are
common to SU UMa type dwarf novae in showing superoutbursts and superhumps.
Remarkable properties of WZ Sge type are an extremely long 
recurrence time ($\sim $ 10 yr), a very large 
outburst amplitude (exceeding 6 mag), a long duration ($\sim$ a month or more)
and lack of normal outbursts.
Early superhumps are observed only in early
stages of superoutbursts of WZ Sge type dwarf novae.
Early superhumps are doubly-waved humps with a period nearly equal to the
 orbital period 
(\cite{Bohusz1979}, \cite{Patterson1981}, \cite{Kato1996}, \cite{Nogami1997}, \cite{Matsumoto1998}, \cite{Kato2001},\cite{Ishioka2001}, \cite{Ishioka2002}, \cite{Patterson2002}).
Their amplitudes are smaller than 0.2 mag. 
Early superhumps of WZ Sge show a small bump at orbital phase of 0.1-0.2,
a small dip at 0.3-0.4 and large bump at 0.6-0.7 \citep{Ishioka2002}.
\citet{Osaki2002} proposed that early superhumps are caused by two-armed 
dissipation pattern on the accretion disk \citep{Lin1979}.
In dwarf novae with an extremely small mass ratio, 
the accretion disk can 
expand beyond the 2:1 resonance radius when an outburst occurs. Strong 
two-armed dissipation pattern appears due to the 2:1 resonance, 
and doubly-waved periodic modulations are observed.
The another model of early superhumps is proposed by \citet{Kato2002}.This model
 is an application of the tidal distortion effect on
 accretion disks (\cite{Smak2001}; \cite{Ogilvie2002}).
\citet{Kato2002} proposed that both the two-armed spiral structures in
 Doppler maps and early superhumps can be
 explained by irradiation of an elevated accretion disk due to the tidal
 distortion effect.

BC UMa was discovered by \citet{Romano1964} as a dwarf nova with an
 outburst amplitude of 7 mag.
\citet{Howell1990} detected humps with a period of 91 min
and an amplitude of 0.25 mag in the quiescence phase.
A spectrum taken by \citet{Mukai1990} during a faint state shows both
 absorption and emission components in its Balmer lines and TiO band 
feature at 760 nm.
\citet{Mukai1990} estimated several system parameters as below:  
the spectral type of the secondary is later than M5,
the distance is 130-400 pc, 
and the absolute magnitude ($M_V$) is 11.0-13.5 mag.
\citet{Patterson2003} carried out time-resolved photometry during 
the 2000 April superoutburst.
They detected early superhumps with a period of 0.06256(8) d during the 
first 4 days of its superoutburst as well as superhumps
 with a period of 0.06452(9) d after that.
They also obtained the radial velocity curve and determined the orbital period
 to be 0.06261(4) d. 
The period of the early superhumps is equal to orbital period within the errors.
The superhump period excess is 3\%.

\citet{Schmeer2003} reported an outburst of BC UMa at 2003 February 1.205(UT).
We started time-resolved photomemory on February 1.603(UT), 9 hours after
 we received the outburst report.
In this paper we report our observations during the superoutburst in February 
2003 and we have further modelled light curves of early superhumps.

\section{Observation}
We observed BC UMa at Saitama and Mie in Japan.
Our observation logs and instruments are summarized in table\ref{logobs}.
All frames were dark-subtracted and flat-fielded before photometry.
The Saitama frames were processed by the aperture photometry package in IRAF
\footnote{
IRAF is distributed by the National Optical Astronomy Observatories for Research in Astronomy Inc. under cooperative agreement with the National Science Foundation.
}.
The frames obtained at Mie were processed by the aperture photometry packages, FITS Photo, developed by K.Nagai
\footnote{FITS Photo is an aperture photometry software developed by Kazuo Nagai. This software is available at 
http://www.geocities.jp/nagai\_kazuo/index-e.html.}.
The magnitudes were measured by using the local standard stars,
TYC2 3454.875.1(Mie) and GSC 3454.0868(Saitama).
The constancy of brightness of the comparison stars during the run was checked by using  GSC 3454.0817.

Because each observer used different filters and comparison stars, all data were adjusted to match
$V$-band magnitude obtained at Mie.
The barycentric correction was applied to the observation time before analysis.

\begin{longtable}{rcccc}
\caption{Log of observations.}
\label{logobs}
\hline
\hline
Start-End (UT) & Exposure time(s) & Flame number & Filter & Instrument \\
\endfirsthead
\hline
\hline
Start-End (UT) & Exposure time(s) & Flame number & Filter & Instrument \\
\hline
\hline
\endhead
\hline
\hline
\endfoot
\hline
\multicolumn{5}{l}{
\small{
S: 20cm Newtonian + SBIG ST-7E (Saitama, Japan),
}
}\\
\multicolumn{5}{l}{
\small{
M: 25cm SCT + MUTOH CV-04(Mie, Japan)
}
}
\endlastfoot
\hline
2003 Feb. 1.603-1.822 & 40 & 213 & no & S\\
1.620-1.868 & 60 & 163 & V & M\\
2.542-2.867 & 45 & 322 & V & M\\
3.543-3.799 & 30 & 299 & V & M\\
3.694-3.837 & 40 & 143 & no & S\\
5.569-5.869 & 30 & 305 & V & M \\
6.576-6.873 & 30 & 399 & V & M\\
6.743-6.870 & 40 & 138 & no & S\\
9.591-9.826 & 30 & 210 & V &M\\
9.617-9.661 & 40 & 61 & no & S\\
11.768-11.865 & 30 & 118 & V & M\\
12.551-12.858 & 30 & 400 & no & M\\
\end{longtable}

\section{Results}
\subsection{Early superhumps}
\begin{figure}
\begin{center}
\FigureFile(80mm,50mm){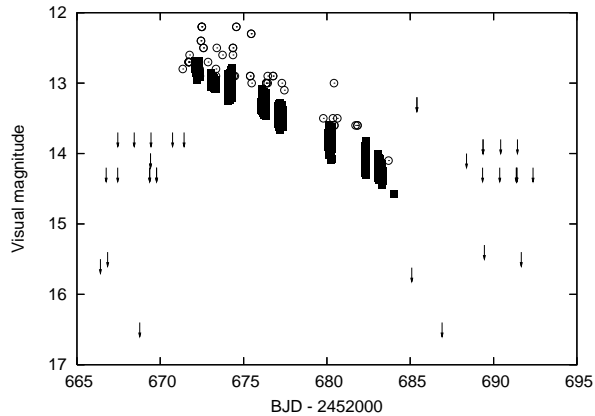}
\end{center}
\caption{Long-term light curve of the 2003 February superoutburst. Filled-squares are our CCD observations. Open circles are visual observations and the downward arrows are the upper limits taken from AAVSO, VSOLJ, and VSNET.}
\label{longterm}
\end{figure}

\begin{figure}
\begin{center}
\FigureFile(80mm,50mm){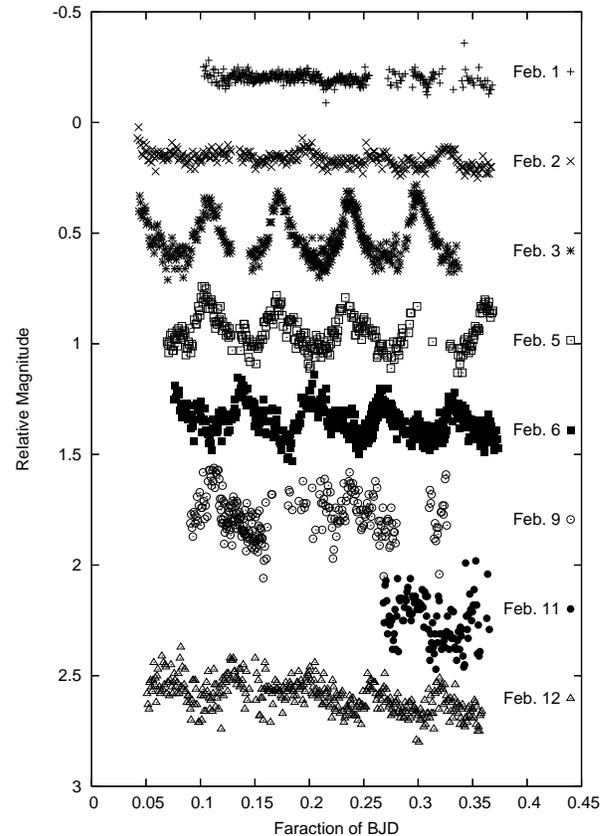}
\end{center}
\caption{Daily light curves of the 2003 February superoutburst. Each daily dataset is shifted as follow: -0.2 on Feb. 1, 0.0 on Feb. 2, +0.3 on Feb. 3, +0.5 on Feb. 5, +0.7 on Feb. 6, +0.8 on Feb. 9, +1.0 on Feb. 11, +1.2 on Feb. 12.}
\label{daily}
\end{figure}

\begin{figure}
\begin{center}
\FigureFile(80mm,50mm){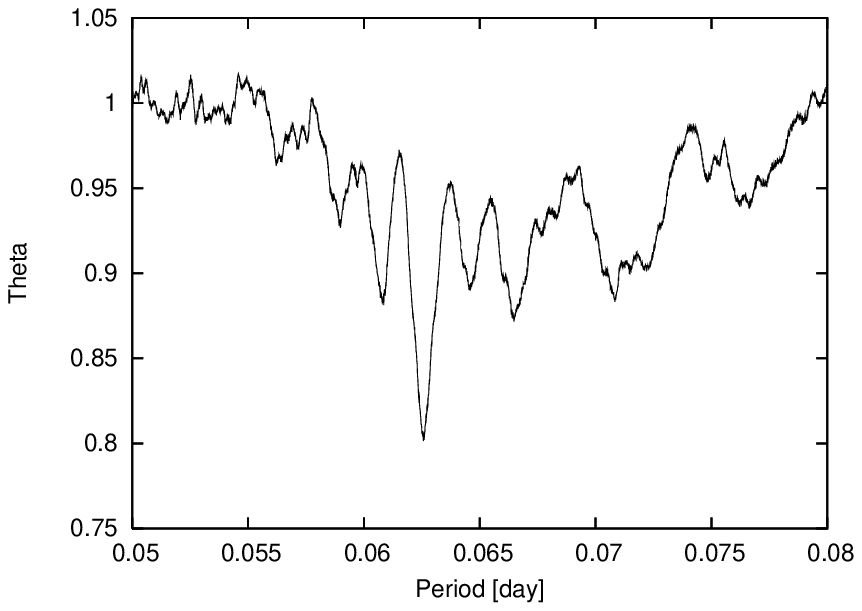}
\end{center}
\caption{Period - theta diagram obtained by the PDM period analysis for the data between February 1 and 2.}
\label{pdm_early}
\end{figure}

\begin{figure}
\begin{center}
\FigureFile(80mm,50mm){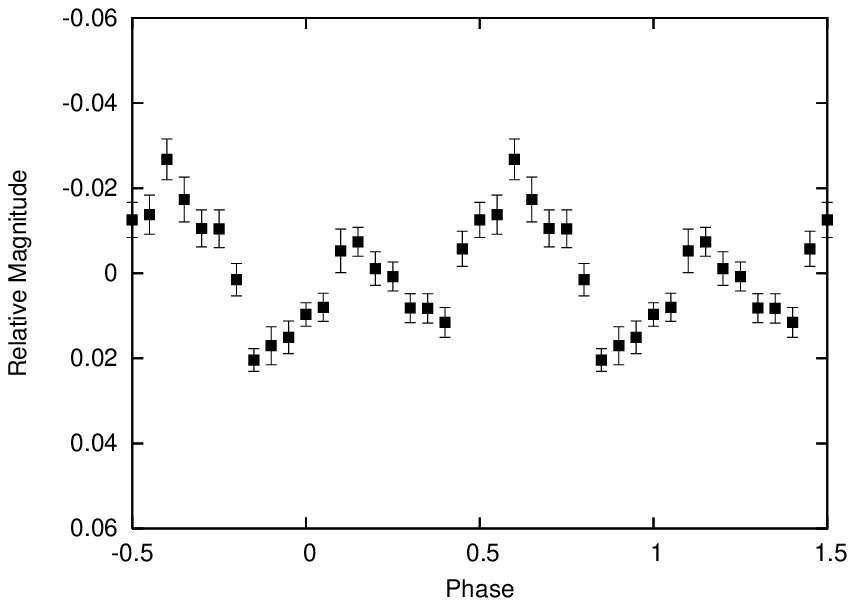}
\end{center}
\caption{Phase-averaged light curve of early superhumps. Since the exact binary phase is  unknown for BC UMa, the phase was taken arbitrarily so that the early superhump maximum phase fit to that of Fig.7 of \citet{Patterson2003}.}
\label{earlyhump}
\end{figure}

Fig. \ref{longterm} shows the long-term light curve of the 2003 February superoutburst.
According to the AAVSO, VSOLJ, and VSNET data, BC UMa was fainter than 13.7 mag 
7 hours before the outburst detection and 16 hours before we have started the
time-series photometry. This suggests that our observations cover the early
 phase of the superoutburst.
Daily light curves are depicted in Fig. \ref{daily}.
On the first and second nights(February 1 and 2) , the object showed small($\Delta m_V \sim 0.04$) variation.
We applied the Phase Dispersion Minimization(PDM) method \citep{Stellingwerf1978} to 
the data which obtained on February 1 and 2 after substraction of the slow decay trend.
The resultant period - theta diagram is shown in Fig. \ref{pdm_early}.
Fig. \ref{earlyhump} shows the phase-averaged light curve of the small 
amplitude modulations during the early phase of superoutburst.
The shape of the modulations is double-peaked and each peak in an orbital cycle
has different amplitudes. These properties are similar to that of  
early superhumps in WZ Sge-type dwarf novae (e.g. \cite{Ishioka2002}). 
The best estimated period of this variation is 0.06258$\pm $0.00013 d.
The error was estimated by the Lafler-Kinman method \citep{Fernie1989}.
This period is equal to the orbital period (0.062605$\pm $0.000011 d:
 \cite{Patterson2003}) within the errors.
The early superhump period of known WZ Sge-type dwarf novae is nearly
equal to the orbital period (e.g. \cite{Ishioka2002}).
Thus we concluded that these modulations are the early superhumps.
The best estimated early superhump period of the 2003 February superoutburst
is equal to that of the 2000 April superoutburst
(0.06256$\pm$0.00008 d: \cite{Patterson2003}) within the errors.

\citet{Ishioka2002} reported that the early superhump period of WZ Sge itself
is slightly shorter ($\sim $0.05\%) than the orbital period.
Because of low accuracy of the early superhump period, we could not confirm the
difference between the orbital period and the early superhump period.

\subsection{Superhumps}
\begin{figure}
\begin{center}
\FigureFile(80mm,50mm){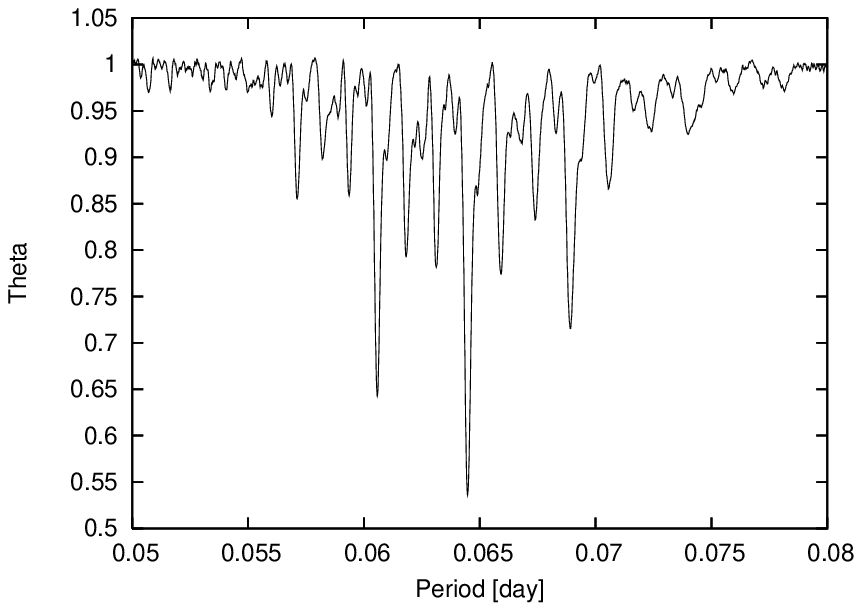}
\end{center}
\caption{Period - theta diagram obtained by the PDM period analysis for the data between February 3 and 12.}
\label{pdm_super}
\end{figure}

\begin{figure}
\begin{center}
\FigureFile(80mm,50mm){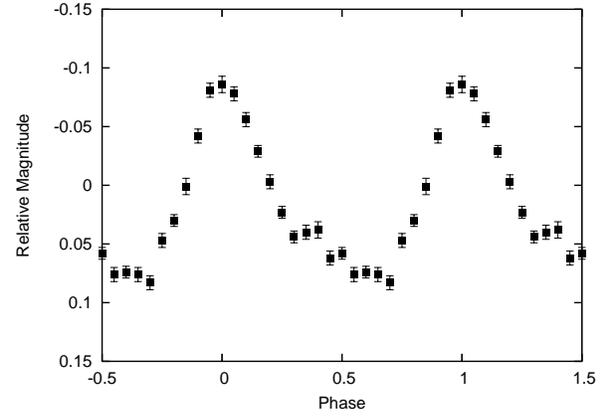}
\end{center}
\caption{Phase-averaged light curve of common superhumps. The superhump phase was 
defined as superhump maximum at phase 0.}
\label{superhump}
\end{figure}

From the third night(February 3), common superhumps clearly appeared.
On February 3 (BJD2452674), BC UMa showed small brightening.
In EG Cnc,  the same kind of brightening was also observed
 (\cite{Matsumoto1998}, \cite{Patterson1998}).
These behavior suggest development of the 3:1 resonance.
After removing the daily trend of brightening and decline, we also
analyzed the data between 2003 February 3 and 12 by the PDM method.
Fig. \ref{pdm_super} shows the resultant period-theta diagram.
The best estimated superhump period($P_{\rm sh}$) of $P_{\rm sh} = 0.06448\pm 0.00006$ d  is 
equal to that in the 2000 February superoutburst ( $P_{\rm sh} = 0.06452\pm0.00009$ d: \cite{Patterson2003}) within the errors.
The superhump period excess is $3.0\pm 0.1$\%.
Fig. \ref{superhump} shows the phase-averaged light curve of superhumps
from the data during the superoutburst plateau phase.
The mean amplitude of the common superhumps was $\Delta m_V\sim 0.2$.

\subsection{Superhump period change}
We estimated superhump maximums by the following way:

\begin{enumerate}
\item We determine nightly "model superhump light curve" from the
 phase averaged superhump light curve of each night.
\item  We shift the maximum timing of the model light curve around 
the eye estimated maximum in a 0.0001 days step, and calculate each
 sum of square of difference between the model light curve and the observation 
for each test maximum timings.
\item The maximum timing is determined by minimizing the sum of square of
residuals divided by the number of data used to calculate the sum. 
\end{enumerate}
The errors of maximum timings were determined to satisfy a 99\% confidence level.

\begin{figure}
\begin{center}
\FigureFile(80mm,50mm){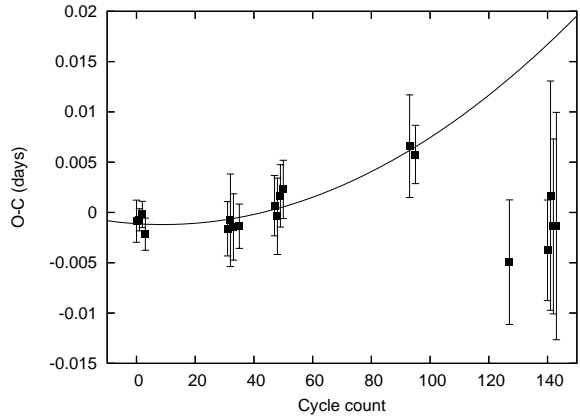}
\end{center}
\caption{$O-C$ diagram for the superhump maxima during the 2003 February superoutburst. The solid curve corresponds to equation (\ref{eq:o-c}).}
\label{o-c_grp}
\end{figure}

The times of superhump maxima are given in table \ref{o-c}.
The cycle count ($E$) is defined as $E=0$ at the first superhump maximum
we observed.
A linear ephemeris of the superhump maximum timings is given by
\begin{equation}
{\rm BJD(maximum)} = 2452674.1078(11)+0.064463(13)E \label{eq:superhump_max}.
\end{equation}
Fig. \ref{o-c_grp} shows the $O-C$ diagram for these superhump maximum timings.
The $O-C$s between $E=0$(February 3) and $E=95$(February 9) can be well represented by 
\begin{eqnarray}
O-C({\rm days}) = & 1.04(\pm 0.26)\times 10^{-6} E^2 - 1.9(\pm 2.5)\times 10^{-5}E\nonumber\\
      & + 9.6(\pm 5.1)\times 10^{-4}\label{eq:o-c}.
\end{eqnarray}
The quadratic term corresponds to $\dot{P_{\rm sh}}/P_{\rm sh} = (+3.2\pm 0.8)\times 10^{-5}\ {\rm [cycle\ count}^{-1}{\rm ]}$.

\begin{table}
\begin{center}
\caption{Times of superhump maxima}
\label{o-c}
\begin{tabular}{rccc}
\hline
\hline
BJD - 2450000 & Error & $E$ & $O-C$\footnotemark[$*$]\\\hline
2674.1061 & 0.0021 & 0 &  -0.0007\\
2674.1717 & 0.0011 & 1 &  -0.0006\\
2674.2367 & 0.0013 & 2 &   0.0000\\
2674.2992 & 0.0016 & 3 &  -0.0020\\
2676.1047 & 0.0027 & 31 & -0.0015\\
2676.1700 & 0.0046 & 32 & -0.0006\\
2676.2338 & 0.0033 & 33 & -0.0013\\
2676.3628 & 0.0022 & 35 & -0.0012\\
2677.1384 & 0.0030 & 47 &  0.0008\\
2677.2018 & 0.0038 & 48 & -0.0003\\
2677.2683 & 0.0031 & 49 &  0.0018\\
2677.3334 & 0.0029 & 50 &  0.0024\\
2680.1096 & 0.0051 & 93 & 0.0067\\
2680.2377 & 0.0029 & 95 & 0.0059\\
2682.2898 & 0.0062 & 127 & -0.0049\\
2683.1290 & 0.0050 & 140 & -0.0037\\
2683.1989 & 0.0114 & 141 &  0.0018\\
2683.2603 & 0.0087 & 142 & -0.0013\\
2683.3248 & 0.0113 & 143 & -0.0013\\
\hline
\multicolumn{4}{@{}l@{}}{\hbox to 0pt{\parbox{85mm}{\footnotesize
\footnotemark[$*$] Using equation (\ref{eq:superhump_max}).
}\hss}}
\end{tabular}
\end{center}
\end{table}

\section{Numerical simulation of early superhumps}
The superhump excess of BC UMa is  larger than that of WZ Sge ($\varepsilon = 1.0$\% : \cite{Ishioka2002}). This suggests that the mass ratio of BC UMa is larger than that of WZ Sge.
In this section, first, we will show how to construct the binary model that reproduce
observed early superhumps, and then, we will present our numerical results.

\subsection{Light curve model of early superhumps}
It is well known that there is a relation between the superhump excess and
 the mass ratio \citep{Osaki1985}. 
We estimated the mass ratio from the superhump excess
using the following equation \citep{Patterson1998}:
\begin{equation}
\varepsilon = \frac{q}{(1.1q+4.3)}\ .
\end{equation}
The superhump excess of BC UMa is 3.0\% and this corresponds to $q=0.13$.
This mass ratio is twice the mass ratio of WZ Sge
 ($q=0.06$, \cite{Skidmore2002}), and 
as large as that  of RZ Leo ($q=0.14$, \cite{Ishioka2001}).

\citet{Osaki2002} proposed that the light source of early superhumps in a
dwarf nova with an extremely low mass ratio like WZ Sge is the two armed
dissipation pattern by the 2:1 resonance. But in dwarf novae with $q>0.08$, the
accretion disk cannot extend to the 2:1 resonance radius. So we assume the
light source of early superhumps in BC UMa is two armed spiral shocks
 on the accretion disk(e.g. \cite{Makita2000}).

\citet{Mukai1990} reported that the spectral type of the secondary star in
 BC UMa is later than M5. 
So we assume a 0.1$\MO$ secondary star with a
surface temperature of 3000K.
Our adopted values for the mass and surface temperature of the secondary star are 
consistent with the results of \citet{Ciardi1998}.
Then the mass of the white dwarf is 0.77$\MO$ from the mass ratio
 of 0.13.
The surface temperature of the white dwarf in an SU UMa-type star Z Cha is
$T_{eff}=17400$K immediately after the outburst \citep{Wood1993}.
\citet{Gansicke2005} determine the effective temperature of the white 
dwarf in BC UMa to be 15200$\pm$1000 K
 at around 100 days after the superoutburst.
So we assume the surface temperature of the white dwarf to be 15000 K.

A circular orbit is assumed.
We also assume the surfaces of the white dwarf, secondary star and accretion
disk emit photons as a blackbody at a local temperature.
The size and thickness of accretion disk are assumed as follows:
\begin{equation}
\frac{R_{\rm disk}}{a} = \left(\frac{7}{5}\right)^2(1+q)\left(\frac{R_{\rm L1}}{a}\right)^4 ,
\end{equation}
\citep{Osaki2002}, and
\begin{equation}
h=\beta R_{\rm disk}\left(\frac{r}{R_{\rm disk}}\right)^\nu,
\end{equation}
where $R_{\rm disk}$ is the radius of the accretion disk, $a$ is the 
binary separation, $q$ is the mass ratio($M_2/M_1$), $R_{\rm L1}$ is
 the distance from the center of the white dwarf to the inner Lagrangian point, 
$h$ is the height of 
accretion disk surface from equatorial plane, and $r$ is the distance from
 the center of the white dwarf.
In the case of $q=0.13$, the accretion disk radius is $0.52a$.
We adopt $\nu = 2$.

\begin{figure}
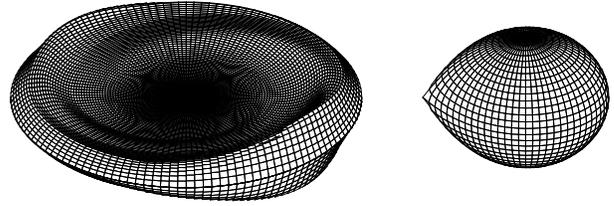

\begin{center}
\FigureFile(80mm,50mm){figure08.eps}
\end{center}
\caption{
Geometrical configurations of our BC UMa model at the binary phase 0.25.
}
\label{bin_model}
\end{figure}

Spiral structures of surface of the accretion disk are assumed in the
same manner as that adopted by \citet{Hachisu2004}, which is defined by 
\begin{equation}
z_1=\max\left(1,\frac{\xi _1}{\sqrt{\{r/R_{\rm disk}-\exp[-\eta(\phi - \delta)]\}^2+\epsilon ^2}}\right),
\end{equation}
\begin{equation}
z_2=\max\left(1,\frac{\xi _2}{\sqrt{\{r/R_{\rm disk}-\exp[-\eta(\phi - \delta - \pi)]\}^2+\epsilon ^2}}\right),
\end{equation}
\begin{equation}
h'=h \max(z_1,z_2),
\end{equation}
where $h'$ is the height of accretion disk surface including the axisymmetric
 structure.
The various disk parameters above are assumed to be $\epsilon = 0.1$,
$\xi _1 = 0.25$,$\xi _2 = 0.15$, $\eta = 0.2$, and $\delta = 110^\circ$ for Fig. \ref{bin_model}.

The nonirradiated surface temperature is one determined by the
viscous heating of the standard disk model \citep{Shakura1973}.
The mass accretion rate is $1\times 10^{-9}\MO/{\rm yr}$
\citep{Horne1985}.
The outer rim of the disk is not irradiated by the white dwarf.
The brightness temperature of the outer rim of the disk in an SU UMa-type
star Z Cha is $\log T \sim 3.9$ at 0.7 days after maximum 
\citep{Horne1985}.
We assumed the temperature of the outer rim of the disk to be 8000 K.

The surfaces of the white dwarf, the accretion disk, and the secondary star
are divided into patches as shown in Fig. \ref{bin_model}. 
We assume that each patch emits photons as a single temperature blackbody.
The patches of the disk and the secondary are irradiated by the front side 
patches of the white dwarf if there is no patch between them.
The total luminosity  of the binary system is calculated by summing up 
each luminosity from all visible patches.
The detail of the numerical method adopted here was described in \authorcite{Hachisu2001}(\yearcite{Hachisu2001},\yearcite{Hachisu2003a},\yearcite{Hachisu2003b},\yearcite{Hachisu2003c}).
The surface patch elements are 32$\times$64 ($\theta \times \phi$) for the
 secondary , $64\times 128\times 2$ ($\theta \times \phi \times$ up 
and down side) for the accretion disk, and $16\times 32$ for the white dwarf.
The number of total time steps are 128 for one orbital period.

\subsection{Numerical results}
\begin{figure}
\begin{center}
\FigureFile(80mm,50mm){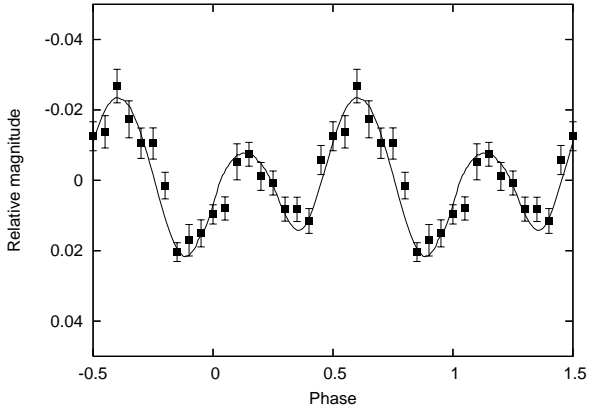}
\end{center}
\caption{Calculated $V$ light curve plotted against the binary phase
 together with the observational points.}
\label{model_lc}
\end{figure}

The best-fit light curve model is plotted in Fig. \ref{model_lc}
together with our observational points (same as Fig. \ref{earlyhump}).
We have changed the inclination angle ($i$), the thickness of the accretion
disk ($\beta$),
 and four parameters of the asymmetric structure of the accretion disk
($ \delta, \eta, \xi _1, \xi _2$) as shown in table \ref{params}.
$\xi _1$ and $\xi _2$ are parameters for the enhancement of thickness on
 the spiral structures.
$\delta$ means the position angle of the spiral structure and $\eta$ is
a parameter of tightness of the spiral (large $\eta$ means the loose spiral
and small $\eta$ means the tight spiral).
These are the best set of parameters among those in table 3:
$i=60^\circ,
\beta=0.13,
\xi _1=0.25,
\xi _2 = 0.15,
\delta = 110 ^\circ,$ and
$\eta=0.2$.
The spectrum of BC UMa shows doubly-peaked Balmer emissions 
\citep{Patterson2003}. This indicates that the inclination is not low.
But absence of eclipse in photometric data suggests $i < 70 ^\circ$.
The best estimated inclination is also consistent with these observational
 results.

The $i$ and $\beta$ affect the total brightness of binary system
and the amplitude of orbital modulations. 
$\xi _1$ and $\xi _2$ affect the amplitude of two bumps of the early superhump.
$\delta$ affects the phases of bumps. 
$\eta$ affects the shape of early superhumps, especially at large $\beta$.
At $\delta = 110 ^\circ$, $\xi _1$ affects the amplitude of the bump around
phase 0.6 and $\xi _2$ affects the amplitude of the bump around phase 0.1.
$\xi _1$, $\xi _2$ and $\eta$ also affect
the total brightness. But the total brightness is more sensitive to $i$ and $\beta$ than to $\xi _1$, $\xi _2$ and $\eta$.
Thus the distance to BC UMa are mainly affected by the inclination angle
 and the thickness of the accretion disk.
By comparing the apparent magnitudes of BC UMa during the early superhump phase
 with the calculated $V$ magnitudes, we have estimated the distance of
BC UMa to be $d = 270 \pm 20 $pc (this error does not include the ambiguity 
of our model itself).

\begin{longtable}{rc}
\caption{Tested parameters}
\label{params}
\hline
\hline
\endfirsthead
\endhead
\hline
\endfoot
\hline
\endlastfoot
$i$ & 45$^\circ$, 50$^\circ$, 55$^\circ$, {\bf 60$^\circ$}, 65$^\circ$, 70$^\circ$, 75$^\circ$\\
\hline
$\beta$ & 0.05, 0.07, 0.09 , 0.11, {\bf 0.13}, 0.15\\
$\eta$ & 0.1, {\bf 0.2}, 0.4\\
$\delta$ & 10$^\circ$, 20$^\circ$, 30$^\circ$, 40$^\circ$, 50$^\circ$, 60$^\circ$, 70$^\circ$, 80$^\circ$, 90$^\circ$, 100$^\circ$, {\bf 110$^\circ$}, 120$^\circ$, 130$^\circ$, 140$^\circ$, 150$^\circ$, 160$^\circ$, 170$^\circ$\\
$\xi_1 $ & 0.15, 0.20, {\bf 0.25}, 0.30\\
$\xi_2 $ & 0.10, {\bf 0.15}, 0.20, 0.25\\
\end{longtable}

\section{Discussion}
\subsection{Distance}
In this subsection, we estimate the distance to BC UMa based on the maximum
 magnitude of normal outbursts. 
We analyzed the data of the AAVSO International Database, VSOLJ
 and VSNET and found some normal outbursts.

\citet{Warner1987} proposed an empirical relationship between the
 absolute magnitudes at the normal outburst maximum and the orbital periods, i.e.,
\begin{equation}
	M_{V}({\rm max})=5.74 - 0.259 P_{{\rm orb}} ({\rm h}).
\end{equation}
To estimate the correct maximum magnitude of normal outbursts, we searched for
outburst observations which satisfy the following conditions:

\begin{description}
	\item[C1] The duration of an outburst was a few days.
	\item[C2] Within one day before the outburst was detected,
some observers checked that BC UMa was in faint state.
	\item[C3] More than two observers detected the outburst.
\end{description}
The first condition C1 is the definition of normal outburst \citep{Warner1985},
 and the second one C2 is used to determine the true maximum magnitude.
Because most of data are visual observations, we pose the third condition C3
 to exclude the misidentification or any other errors.

We found two normal outbursts which satisfy all the conditions above 
and one normal outburst which falls short of the second condition
as shown in table \ref{outbursts}.
The maximum magnitude of normal outbursts is $13.1\pm 0.1$ (the error is the typical error for visual observations).
The orbital period of BC UMa is $0.062605\pm 0.000011$ d \citep{Patterson1998},
 corresponding to $M_V = 5.35\pm 0.23$. 
Following \authorcite{Warner1987}'s(\yearcite{Warner1987}) equation
we assume the inclination effect 
\begin{equation}
\Delta M_V(i)=-2.5\log\left\{\left(1+\frac{3}{2}\cos i\right)\cos i\right\}.
\end{equation}
The inclination angle of the best-fit numerical model of 
early superhumps ($i=60^\circ$) corresponds to $\Delta M_V = 0.145$.
 Thus the absolute magnitude of BC UMa is $M_V = 5.50 \pm 0.23$.
Assuming the apparent maximum magnitude, $m_V = 13.1\pm 0.1$, 
the distance is calculated to be $330\pm 50$ pc.
This value is consistent with
 the distance estimated by our numerical model ($270\pm 20$ pc)
and \citet{Gansicke2005}, value of 285$\pm$ 42 pc
from the FUV flux within the errors.

\begin{longtable}{rcccc}
\caption{Normal outbursts and superoutbursts observed by AAVSO, VSOLJ, and VSNET.}
\label{outbursts}
\hline
\hline
Date\footnotemark[*] & JD& duration(days) & maximum magnitude & type\\
\endfirsthead
\hline
\hline
Date\footnotemark[*] & JD& duration(days) & maximum magnitude & type\\
\hline
\hline
\endhead
\hline
\hline
\endfoot
\hline
\multicolumn{4}{@{}l@{}}{\hbox to 0pt{\parbox{85mm}{\footnotesize
\par\noindent
\footnotemark[*]{The discovery date}
\par\noindent
\footnotemark[$\dagger$]
{Because of no observations within one day before the outburst detection
or the end of outburst,we cannot determine the accurate outburst duration.}
}\hss}}
\endlastfoot
\hline
1982 May 20 & 2445109 & 21-22\footnotemark[$\dagger$] & 11.0 & super \\
1987 May  8 & 2446924 & 4-7\footnotemark[$\dagger$] & 13.1 & normal \\
1990 Sep. 9 & 2448143 & 15-19\footnotemark[$\dagger$] & 10.9 & super \\
1992 Aug. 18 & 2448852 & 9-13\footnotemark[$\dagger$] & 12.5 & super \\
1994 Apr. 27 & 2449469 & 16-17\footnotemark[$\dagger$] & 12.0 & super \\
1995 Jul. 6 & 2449904 & 3  & 13.0 & normal \\
2000 Mar. 31 & 2451635 & 23 & 11.5 & super \\ 
2001 Jun. 25 & 2452085 & 5  & 13.1 & normal \\
2003 Jan. 31 & 2452671 & 11 & 12.6 & super \\
\end{longtable}

\subsection{Supercycle, duration and amplitude of superoutburst}
In this subsection, we estimate the recurrence time of superoutbursts, 
the durations and amplitudes of superoutbursts
from the data of AAVSO, VSOLJ and VSNET, and we describe our 
discovery of two types of superoutbursts in BC UMa.

We searched for outburst observations which satisfy the last condition C3 in
 the previous section.
Table \ref{outbursts} shows normal and superoutburst observations in  AAVSO,
VSOLJ, and VSNET data.
The supercycle is from 600 days (no normal outbursts between superoutbursts) 
to 1000 days (one normal outburst between superoutbursts).
This value is longer than that of typical SU UMa type star 
(several hundred days),
but shorter than that of typical WZ Sge type dwarf novae (around 10 years).

The AAVSO, VSOLJ and VSNET data show two types of superoutbursts in BC UMa.
One is short and faint superoutbursts whose durations are around 10 days and
 maximum magnitudes are 12.5 mag (JD 2448852, 2452672).
The other is long and bright superoutbursts whose durations are around 20 days
and maximum magnitudes are 11-11.5 mag (JD 2445110, 2448143, 2449469, 2452085).
The magnitude in quiescence of BC UMa is around $V=18.6$ \citep{Howell1990}.
Thus the amplitudes of outbursts of BC UMa are $\Delta m_V\sim 5.5$
 for normal outbursts,
$\Delta m_V\sim 6$ for faint superoutbursts and $\Delta m_V\sim 7.5$ for
 bright superoutbursts.
The amplitude of faint superoutbursts are smaller than the outburst amplitude 
of typical WZ Sge type dwarf novae ($\Delta m_V=7-8$).
The amplitude of bright superoutbursts are as large as that of WZ Sge type stars.
The durations of faint superoutbursts are shorter than that of WZ Sge type
stars. The durations of bright superoutbursts are as long as those of WZ Sge
type stars.
An SU UMa-type dwarf nova LL And also shows two types of superoutbursts:
the short and faint one is the 1993 December superoutburst \citep{Kato2004}
, the long and bright one is the 2004 May superoutburst (\cite{Waagen2004}
and VSNET data).

The 2003 February superoutburst of BC UMa was the short and faint superoutburst.
The short duration and faint maximum magnitude indicate that the
stored mass in the accretion disk before the superoutburst was smaller
than that of WZ Sge type stars.

\citet{Mukai1990} noted that the spectrum of BC UMa is very similar to that
of WZ Sge. 
The presence of normal outbursts and two types of superoutbursts (one has 
 short duration and small amplitude compared with WZ Sge type dwarf novae
and the other has long duration and large amplitude like WZ Sge type stars)
suggest that BC UMa is an intermediate dwarf nova between normal SU UMa type 
 stars and WZ Sge type stars.

\subsection{Brightening phenomenon at the superhump developing phase}
The decline rate and average brightness of nightly data are shown in 
table \ref{declinerate}.
Overall decline rates of the plateau phase are 0.13 mag/day.
This is the typical value for that of SU UMa type dwarf novae.
But on Feb. 3 (BJD2452674), the data indicate a small brightening.
The decline rate before the development of superhumps was nearly equal
to that of plateau phase 
and the brightness before the appearance of superhumps
 was brighter than that after superhumps appeared.
These phenomenon indicate that the small brightening on Feb. 3 was not
the transition phase from the precursor outburst to the main superoutburst.
Thus the 2003 superoutburst was not precursor-main
type superoutburst (\cite{Osaki2003}, \cite{Osaki2005}).

The same phenomenon were observed in EG Cnc
(\cite{Patterson1998}, \cite{Matsumoto1998}) and
V844 Her (\cite{Kato2000}).
By analogy with the mechanism of rising phase of the precursor-main 
type superoutburst (\cite{Osaki2003}, \cite{Osaki2005}),
the small brightening at the superhump developing phase
may be caused by the increased tidal dissipation and tidal torques by the
expanding eccentric disk.

\begin{longtable}{rcc}
\caption{Nightly decline rate and average magnitude during the 2003 superoutburst}
\label{declinerate}
\hline
\hline
Date & decline rate (mag/day) & average magnitude\\
\endfirsthead
\hline
\hline
Date & decline rate (mag/day) & average magnitude\\
\hline
\hline
\endhead
\hline
\hline
\endfoot
\hline
\multicolumn{2}{@{}l@{}}{\hbox to 0pt{\parbox{85mm}{\footnotesize
\par\noindent
\footnotemark[$*$]
The duration of time-series observation are not long enough.
}\hss}}
\endlastfoot
\hline
2003 Feb. 1 & 0.09 & 12.80\\
Feb. 2 & 0.16 & 13.00\\
Feb. 3 & -0.01 & 13.03\\
Feb. 5 & 0.10 & 13.28 \\
Feb. 6 & 0.24 & 13.47 \\
Feb. 9 & 0.03 & 13.80 \\
Feb. 11 & -\footnotemark[$*$] & 14.08\\
Feb. 12 & 0.44 & 14.23\\
\end{longtable}

\subsection{Early superhumps}
Early superhumps in BC UMa were discovered by \citet{Patterson2003} in
the 2000 superoutburst.
We confirmed the existence of early superhumps and its period 
is equal to the orbital period within the errors.
The 2000 superoutburst had long duration (23 days) and large outburst
 amplitude (7 mag). But the 2003 superoutburst had short duration (11 days)
and relatively small outburst amplitude (6 mag). Early superhumps was 
observed in both superoutbursts.

Early superhumps in WZ Sge type dwarf novae are explained by the 2:1 resonance
model \citep{Osaki2002}. 
In dwarf novae with $q < 0.08$, 
the disk expands beyond the 2:1 resonance radius.
A hydrodynamic simulation of the accretion disk in binary system with a low
mass ratio showed that two-armed dissipation patterns appear near 
the 2:1 resonance radius \citep{Lin1979}.
But the mass ratio of BC UMa ($q=0.13$)
is much larger than 0.08 and twice as large as that of WZ Sge
($q=0.06$ : \cite{Skidmore2002}).
This indicates that the disk does not expand beyond the 2:1 resonance radius.

The duration of the outburst is 11 days and the 
outburst amplitude is 6 mag for the 2003 superoutburst of BC UMa.
But WZ Sge type stars have longer outburst durations (around 1 month) and
larger outburst amplitudes (7-8 mag).
The short duration and small outburst amplitude suggest that the stored
mass in the accretion disk of BC UMa is smaller than that of WZ Sge type
 stars.

The 2:1 resonance radius in a dwarf nova with $q=0.13$ is nearly
equal to its inner critical Roche lobe radius. So if the accretion disk
grows over the tidal truncation radius and near the inner critical 
Roche lobe radius during the bright superoutbursts,
the 2:1 resonance may cause the early superhumps.
But because of large mass ratio and small mass of accretion disk,
the 2:1 resonance may not occur in BC UMa during the 2003 superoutburst.

According to the refined TTI model \citep{Osaki2003}, 
the accretion disk expands till the tidal truncation radius 
when a superoutburst with no precursor occurs.
Absence of precursors in our observation for the 2003 superoutburst 
of BC UMa (see Fig.\ref{longterm}) indicates that 
the disk of BC UMa expanded near the tidal truncation radius.
Some numerical simulations of the accretion disk showed that tidally induced
spiral shocks are excited on the accretion disk (e.g. \cite{Makita2000}).
Two-armed shocks develop in binary systems with a wide range of mass ratios
(e.g. \cite{Matsuda1990}).
When the disk expands till the tidal truncation radius, strong tidal force
acts on the disk material and then the spiral shocks appear on the disk.
Our numerical results show that spiral structure on the disk can reproduce
the early superhumps even in the dwarf novae with $q=0.13$.
This indicates that early superhumps in BC UMa is caused by the tidally induced
spiral shocks on the accretion disk.

\subsection{Superhump period change}
The superhump period in SU UMa type dwarf novae generally decreases through
a superoutburst. The decreasing rate is an order of
$\dot{P}_{sh}/P_{sh}\sim 10^{-5}$
 (\cite{Warner1985}; \cite{Patterson1993}).
The superhump period decrease was explained by the shrink of the accretion disk
as a superoutburst progresses \citep{Patterson1993}.
On the other hand, increases of the superhump periods were discovered in
WZ Sge type dwarf novae (e.g. \cite{Nogami1997}) and some SU UMa type 
dwarf novae (e.g. \cite{Semeniuk1997}). The mechanism of
the superhump period increases are not yet well understood.

\citet{Uemura2005} discovered a relation between the presence
of precursor and the value of $\dot{P_{sh}}/P_{sh}$ in TV Crv ($q=0.16$, \cite{Uemura2005}).
TV Crv has two different types of superoutbursts;
one with a precursor (called
type A superoutburst) and the other without (called type B superoutburst).
The period change depends on the presence/absence of a precursor.
When type A superoutburst occurs, the period derivation is almost zero.
On the other hand, in type B superoutburst,  the period derivation 
is positive.
The presence of two different types of superoutbursts can be interpreted with 
the refined thermal-tidal instability model \citep{Osaki2003}.
In the case of type B superoutbursts, the disk has enough mass and can expand beyond
the 3:1 resonance. Hence, the accretion disk still has a large amount of gas 
beyond the 3:1 resonance radius even a few days after the superoutburst
maximum. 
\citet{Uemura2005} proposed that the growth mode of the eccentricity of the 
accretion disk
continues to be excited when the accretion disk remains larger than the 3:1 resonance radius.

Because the mass ratio is $q=0.13$ in BC UMa,
 the tidal truncation radius is enough larger than the
3:1 resonance radius. Thus the accretion disk of BC UMa may expand
beyond the 3:1 resonance.
The 2003 superoutburst of BC UMa was type B superoutburst (no precursor).
The accretion disk still has a large amount of gas beyond the 3:1 resonance.
Thus the eccentric mode continued to be excited, so that the changing rate of 
superhump period was positive.

\bigskip

We are grateful to many AAVSO, VSOLJ and VSNET observers who have reported 
visual and CCD observations.
We are also grateful to an anonymous referee for valuable suggestions.

\end{document}